\documentclass[epj]{svjour}

\usepackage{graphics}
\usepackage{bm}
\usepackage{graphicx}

\usepackage{amsmath,amssymb,amsfonts}
\usepackage{cite}
\usepackage{lineno}
\usepackage{dcolumn}
\usepackage{bm} 
\usepackage[english]{babel} %

\newcommand{\ssst}{\scriptscriptstyle}
\newcommand{\sst}{\scriptstyle}
\newcommand{\dst}{\displaystyle}

\begin{document}

\title{Electromagnetic $KY$ production from the proton \protect\\ in a Regge-plus-resonance approach} 

\author{T. Corthals \inst{1} \and T. Van Cauteren \inst{1} \and J. Ryckebusch \inst{1} \and D.G. Ireland \inst{2}
}                     

\institute{Department of Subatomic and Radiation Physics, Ghent University, Proeftuinstraat 86, B-9000 Gent, Belgium \and University of Glasgow, Glasgow G12 8QQ, United Kingdom}

\date{Received: date / Revised version: date}

\abstract{A Regge-plus-resonance (RPR) description  of the $p(\gamma,K)Y$ and $p(\mathrm{e},\mathrm{e}'K)Y$ processes ($Y=\Lambda,\Sigma^{0,+}$) is presented. The proposed reaction amplitude consists of Regge-trajectory exchanges in the $t$~channel, supplemented with a limited selection of $s$-channel resonance diagrams. The RPR framework contains a considerably smaller number of free parameters than a typical effective-Lagrangian model. Nevertheless, it provides an acceptable overall description of the photo- and electroproduction observables over an extensive photon energy range. It is shown that the electroproduction response functions and polarization observables are particularly useful for fine-tuning both the background and resonance parameters.
}

\maketitle

\section{Introduction}
\label{sec: intro}

Recent measurements performed at the JLab, SPring-8, ELSA and GRAAL facilities have resulted in an extensive set of precise $p(\gamma,K)Y$~\cite{McNabb04,Bradford06,Lawall05,Zegers03,Sumihama05,Graal06,Schum06} and $p(\mathrm{e},\mathrm{e}'K)Y$~\cite{Mohring03,Carman03,Carman06}  data in the few-GeV regime. These experimental achievements have motivated renewed efforts by various theoretical groups. A great deal of attention has been directed towards the develop\-ment of tree-level isobar models, in which the scatte\-ring amplitude is constructed from a selection of lowest-order Feynman diagrams~\cite{AdWr88,WJ92,DaSa96,MaBeHy95,Stijnprc01,Sara05}, as well as to coupled-channels approaches~\cite{US05,MoShklyar05,Diaz05}. While the latter successfully address a number of issues, the multitude of parameters involved constitutes a complicating factor. It is therefore our opinion that ambiguities related to, for example, form factors, gauge-invariance resto\-ration, or parameterization of the background, can be addressed more efficiently at the level of the individual channels. 

In Refs.~\cite{CorthalsL,CorthalsS}, a tree-level effective-field model was developed for $KY$ ($Y = \Lambda, \Sigma^{0,+})$ photoproduction from the proton. It differs from traditional isobar models in its description of the background contribution to the amplitude,
which involves the exchange of a number of kaonic Regge trajectories in the $t$ channel, an approach pioneered by Guidal and Vanderhaeghen~\cite{reg_guidal}. 
To this Regge background, we added a number of resonant contributions. Such a ``Regge-plus-resonance'' (RPR) strategy has the advantage that the background contribution involves only a few parameters, which can be largely constrained against the high-energy data. Furthermore, the use of Regge propagators eliminates the need to introduce strong form factors in the background terms, thus avoiding the gauge-invariance issues plaguing traditional effective-Lagrangian models~\cite{US05}. 

In this work, the RPR prescription from Refs.~\cite{CorthalsL,CorthalsS} is applied to the electroproduction processes $p(\mathrm{e},\mathrm{e}'K^+)\Lambda,$ $\Sigma^0$. It will be demonstrated that the electroproduction response functions are particularly useful for fine-tuning certain model choices which the photoproduction data failed to determine unambiguously.

\section{Constructing the RPR amplitude}
\label{sec: formalism}

\subsection{Background contributions}

Regge theory rests upon the proposition that, at energies where individual resonances can no longer be distinguished, the reaction dynamics are governed by the exchange of entire Regge trajectories rather than of single particles. This high-energy approach applies in particular to the forward or backward angular ranges, corresponding to $t$- or $u$-channel exchanges, respectively. This work focuses on the forward-angle kinematical region which, for electromagnetic $KY$ production, implies the exchange of kaonic trajectories in the $t$ channel. 

An efficient way to model trajectory exchanges involves embedding the Regge formalism into a tree-level effective-field model~\cite{reg_guidal}. The amplitude for $t$-channel exchange of a linear kaon trajectory 
\begin{equation}
\alpha_X(t)=\alpha_{X,0} + \alpha'_X \, (t-m_X^2)\,,
\label{eq: reggetraj}
\end{equation}
with $m_X$ the mass and $\alpha_{X,0}$ the spin of the trajectory's lightest member (or ``first materialization'')~$X$, can be obtained from the standard Feynman amplitude by replacing the Feynman propagator with a Regge one:
\begin{equation}
\frac{1}{t-m_X^2} \hspace{6pt} \xrightarrow{~\quad} \hspace{8pt} \mathcal{P}^X_{Regge}[s,\alpha_X(t)]\,.
\end{equation}
The Regge amplitude can then be written as
\begin{equation}
\mathcal{M}^X_{Regge}(s,t) = \mathcal{P}^X_{Regge}[s,\alpha_X(t)] ~\times~ \beta_X(s,t) \,,
\label{eq: define_reggeprop}
\end{equation}
with $\beta_X(s,t)$ the residue of the original Feynman amplitude,
to be calculated from the interaction Lagran\-gi\-ans at the $\gamma^{(\ast)} K X$ and $p X Y$ vertices. 

In our treatment of $K^+\Lambda$ and $K^+\Sigma^0$ photoproduction \cite{CorthalsL, CorthalsS}, we identified the $K(494)$ and $K^{\ast}(892)$ trajectories as the dominant contributions to the high-energy amplitudes. The corresponding propagators assume the following form~\cite{Donnachie02}: 
\begin{align}
\mathcal{P}^{K(494)}_{Regge}(s,t) =& \left(\frac{\dst s}{\dst s_0}\right)^{\alpha_K(t)}
\frac{1}{\sin\bigl(\pi\alpha_K(t)\bigr)} \nonumber\\
&\times \frac{\pi \alpha'_K}{\Gamma\bigl(1+\alpha_K(t)\bigr)} \ \left\{ \begin{array}{c}
1 \\ e^{-i\pi\alpha_{K}(t)} 
\end{array}\right\} \,, 
\label{eq: reggeprop_K}
\end{align}
\begin{align}
\mathcal{P}^{K^{\ast}(892)}_{Regge}(s,t) =& \left(\frac{\dst s}{\dst s_0}\right)^{\alpha_{K^{\ast}}(t)-1} 
\frac{1}{\sin\bigl(\pi\alpha_{K^{\ast}}(t)\bigr)} \nonumber \\
&\times \frac{\pi \alpha'_{K^{\ast}}}{\Gamma\bigl(\alpha_{K^{\ast}}(t)\bigr)} \ \left\{ \begin{array}{c}
1 \\
e^{-i\pi\alpha_{K^{\ast}}(t)}
\end{array}\right\}\,,\label{eq: reggeprop_Kstar}
\end{align}
with trajectory equations given by $\alpha_{K}(t) = 0.70 \ (t-m_{K}^2)$, $\alpha_{K^{\ast}}(t) = 1 + 0.85 \ (t-m_{K^{\ast}}^2)$~\cite{CorthalsL}. For each of these propagators, a constant~(1) or rotating~($e^{-i\pi\alpha(t)}$) phase can be selected.

It is argued in Ref.~\cite{reg_guidal} that, for the sake of current conservation, the amplitude for the charged-kaon channels should include the electric ($\sim e \overline{N} \gamma_{\mu} N A^{\mu}$) contribution to the $s$-channel Born term:
\begin{align}
&\mathcal{M}_{Regge}\,(\gamma^{(\ast)}\,p \rightarrow K^+ \Lambda,\Sigma^0) = \mathcal{M}_{Regge}^{K^+(494)} + \nonumber \\
&\ \mathcal{M}_{Regge}^{K^{\ast +}(892)} + \mathcal{M}_{Feyn}^{p\ssst,\sst elec} \times \mathcal{P}_{Regge}^{K^+(494)} \times (t-m_{K^+}^2).\label{eq: gauge_recipe}
\end{align}
Eq.~(\ref{eq: gauge_recipe}) applies to electro- as well as photoproduction. It turns out that the measured $Q^2$ behaviour of the $\sigma_L/\sigma_T$ ratio can only be reproduced provided that the same form factor is used at the $\gamma^{\ast} p p$ and $\gamma^{\ast} K^+ K^+$ vertices~\cite{reg_guidal}. The unknown coupling constants and trajectory phases contained in $\mathcal{M}_{Regge}$ are determined from the high-energy $p(\gamma,K^+)Y$ data. 

A monopole electromagnetic form factor is assumed for the $K^+(494)$ and $K^{\ast +}(892)$ trajectories, with cutoff values chosen so as to optimally match the high-$Q^2$ behavior of the $\Lambda$ and $\Sigma^0$ electroproduction data: $\Lambda_{K^+} = \Lambda_{K^{\ast +}} = 1300$ MeV.

\subsection{Resonance contributions}

Although Regge phenomenology is a high-energy tool by construction, the experimental meson production cross sections are observed to exhibit Regge behavior for photon energies as low as 4 GeV. Even in the resonance region, the order of magnitude of the forward-angle pion and kaon electromagnetic production observables is remarkably well reproduced in the Regge model~\cite{reg_guidal}. 

It is evident, though, that a pure background amplitude cannot be expected to account for all aspects of the reaction dynamics. At low energies, the cross sections reflect the presence of individual resonances. These are incorporated into the RPR framework by supplementing the reggeized background with a number of resonant $s$-channel diagrams. For the latter, standard Feynman propagators are assumed, in which the resonances' finite lifetimes are taken into account through the substitution
\begin{equation}
s - m_{R}^2 \longrightarrow ~ s - m_{R}^2 + im_{R}\,\Gamma_R
\end{equation}
in the propagator denominators, with $m_R$ and $\Gamma_R$ the
mass and width of the propagating state ($R=N^{\ast},\Delta^{\ast}$). 

Further, the condition is imposed that the resonance amplitudes vanish
at large values of $\omega_{lab}$. This is accomplished by including a
Gaussian hadronic form factor $F(s)$ at the $KYR$ vertices:
\begin{equation} 
F(s) = \exp \left\{- \frac{(s-m^2_{R})^2}{\Lambda_{res}^4}\right\}\,,
\label{eq: gaussff}
\end{equation}
A single cutoff mass $\Lambda_{res}$ is assumed for all resonances. Along with the resonance couplings, $\Lambda_{res}$ is used as a free parameter when optimizing the model against the resonance-region data. Our motivation for assuming a Gaussian shape is explained in Ref.~\cite{CorthalsL}. 

The leading diagrams contributing to the reaction amplitude are assumed identical for the photo- and electroinduced processes, as are the various model parameters. Instead of employing the standard phenomenological dipole parameterization, we calculate all electromagnetic $N^{\ast}$ and $\Delta^{\ast}$ form factors in the context of the Lorentz-covariant constituent-quark model (CQM) developed by the Bonn group~\cite{bonn}. 

\section{Results for $\bm{p(}\mathrm{\textbf{e}},\,\mathrm{\textbf{e}}\bm{' K^+)Y}$}
\label{sec: results}

In Refs.~\cite{CorthalsL,CorthalsS}, the RPR prescription
was applied to the various $\gamma p \rightarrow KY$ reaction channels. A number of variants of the RPR model were found to provide a comparably good description of the $\Lambda$, $\Sigma^0$ and $\Sigma^+$ photoproduction observables. Their properties are listed in Table~\ref{tab: kl_models}. Incidentally, the parameters of the $K^+\Lambda$ variants from Ref.~\cite{CorthalsL} have been slightly readjusted in order to describe the new singly-polarized GRAAL data~\cite{Graal06}.

The background contribution to the RPR amplitude involves three parameters: one for the $K(494)$ trajectory ($g_{KYp}$) and two for the $K^{\ast}(892)$ one ($G^v_{K^{\ast}}$ and $G^t_{K^{\ast}}$, corresponding to the vector and tensor couplings). In addition, for each trajectory propagator, either a constant (const.) or rotating (rot.) phase may be assumed. As can be appreciated from Table~\ref{tab: kl_models}, in the $K^+\Lambda$ channel two combinations (rot. $K$, rot. $K^{\ast}$/rot. $K$, const. $K^{\ast}$) lead to a comparable quality of agreement between the calculations and the combined high-energy and resonance-region data. Furthermore, for the latter combination, the available photoproduction data do not allow one to determine the sign of $G^t_{K^{\ast}}$. With respect to the quantum numbers of a potential missing $N^{\ast}(1900)$ resonance, both $P_{11}$ and $D_{13}$ emerged as valid candidates. 

\begin{figure}[t]
\begin{center}
\vspace*{1mm}\includegraphics[width=0.5122\columnwidth, clip]{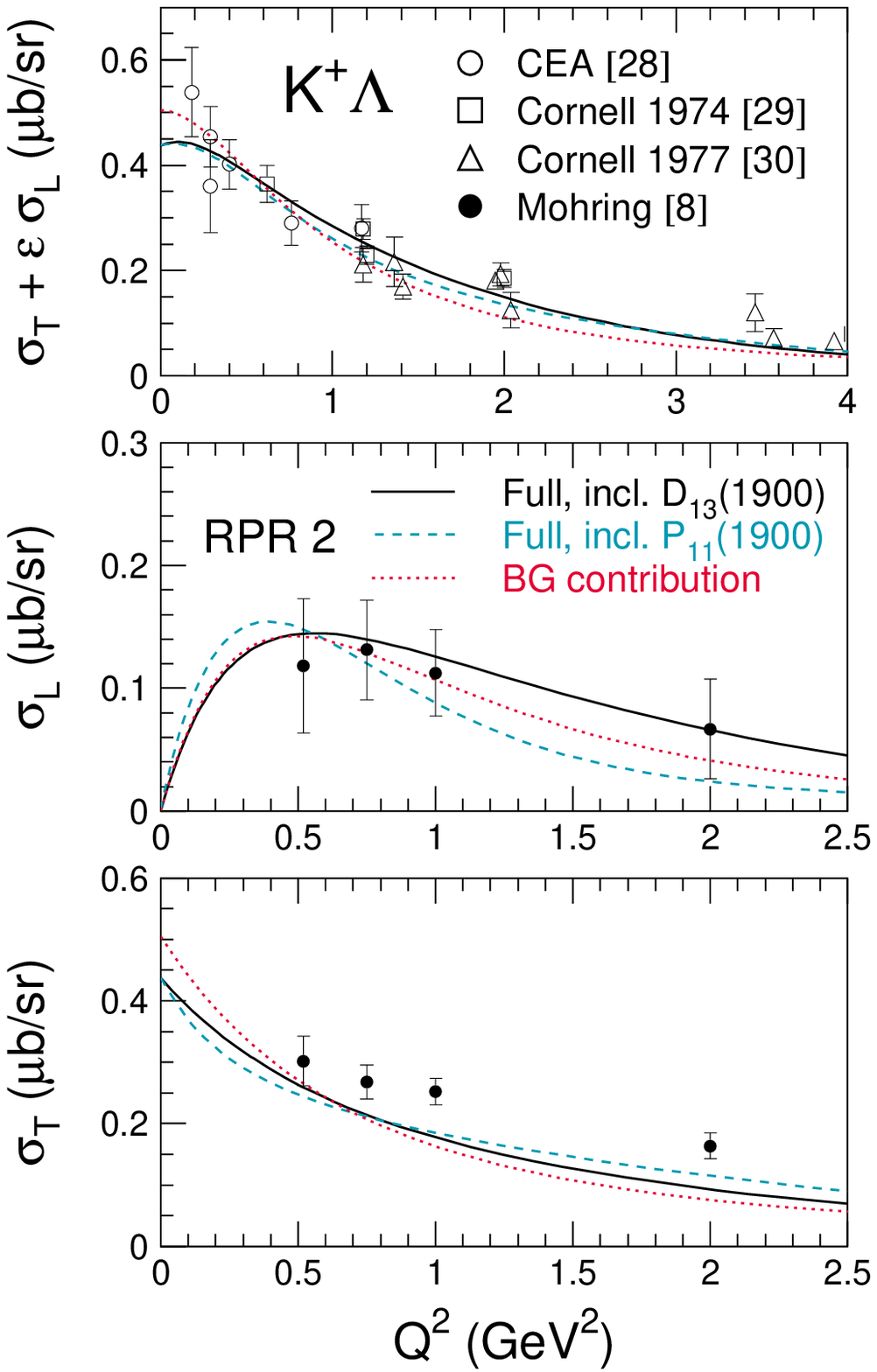}
\includegraphics[width=0.4678\columnwidth, clip]{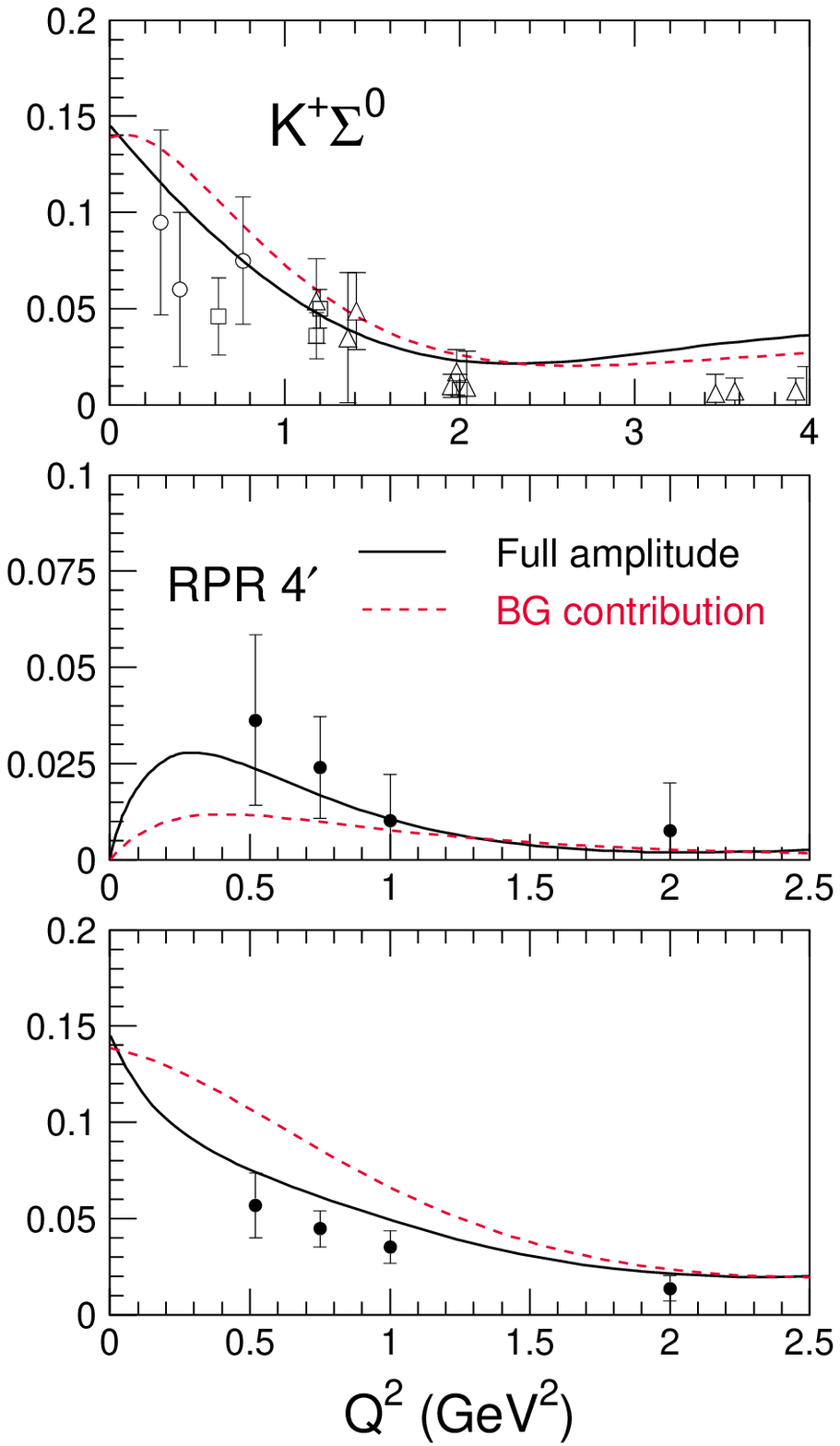}
\caption{$Q^2$ evolution of the unseparated and separated differential cross sections for the $K^+\Lambda$ (left) and $K^+\Sigma^0$ (right) final states. The data for $\sigma_T + \epsilon \sigma_L$ were taken at $W \approx 2.15$~GeV and $\theta_{K}^{\ast} \approx 0$,  and those for $\sigma_L$~and~$\sigma_T$ at $W \approx 1.84$~GeV and $\theta_{K}^{\ast} \approx 8$~deg. For the $K^+\Lambda$ channel, results of the two RPR-2 variants from Table~\ref{tab: kl_models} are displayed, whereas for the $K^+\Sigma^0$ channel, the prediction of the RPR-4$'$ model from Table~\ref{tab: kl_models} is shown. The dotted curves correspond to the Regge background. The data are from~\cite{Brown72,Bebek,Bebek2,Mohring03}.}
\label{fig: k+_lambda_unpol}
\end{center}
\end{figure}

\begin{figure}[t]
\begin{center}
\includegraphics[width=\columnwidth]{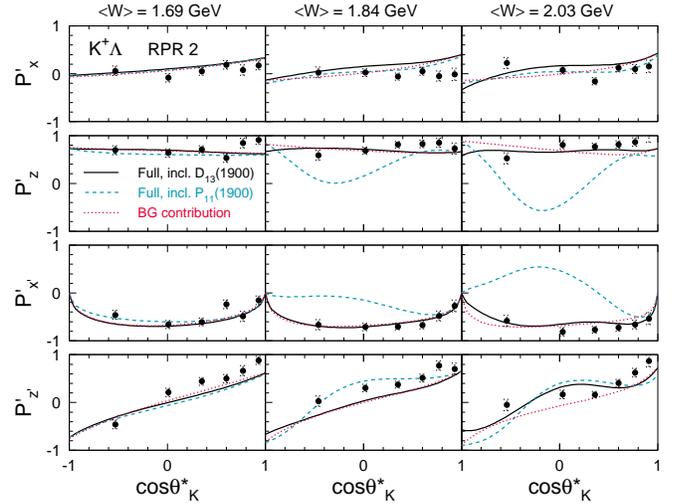}
\caption{Transferred polarization components $P'_x$, $P'_z$, $P'_{x'}$ and $P'_{z'}$ for the $K\Lambda$ final state as a function of $\cos\theta_K^{\ast}$, for $W \approx$ 1.69, 1.84 and 2.03 GeV. Line conventions as in Fig.~\ref{fig: k+_lambda_unpol} (left panel). The data are from \cite{Carman03}.}
\label{fig: k+_lambda_pol}
\end{center}
\end{figure}

\begin{figure}[t]
\begin{center}
\includegraphics[width=0.4233\columnwidth, clip]{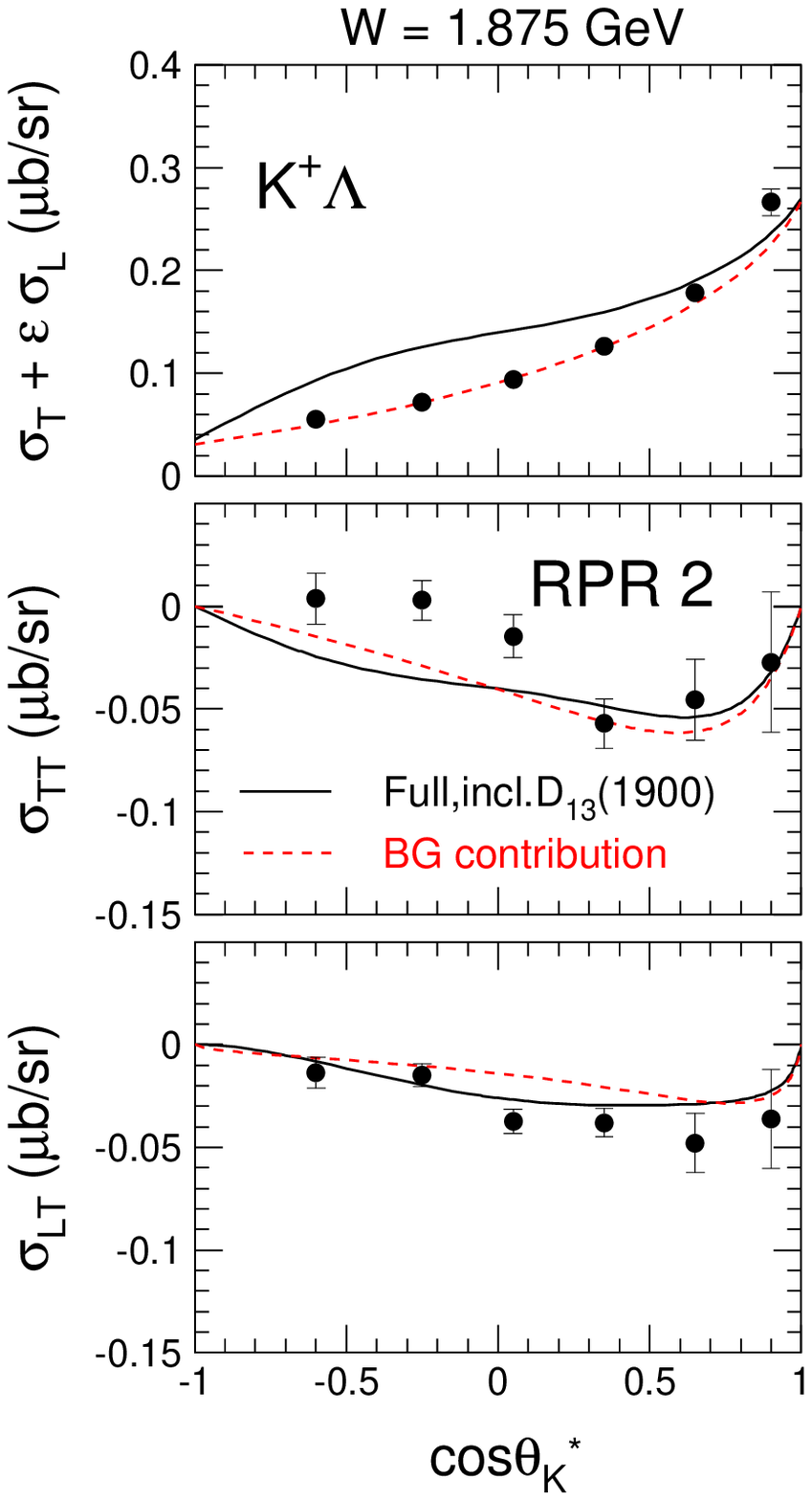}
\includegraphics[width=0.3767\columnwidth, clip]{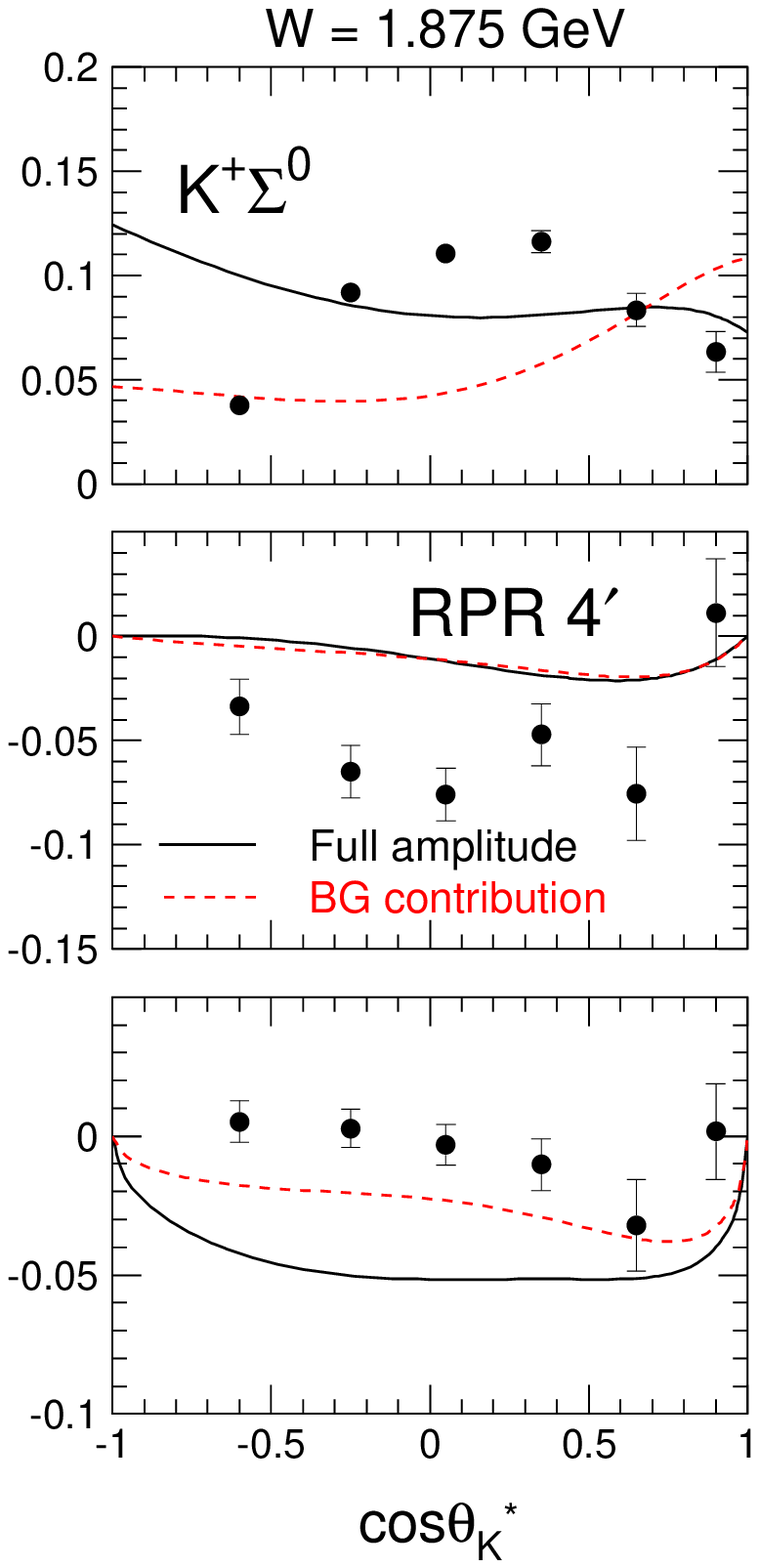}
\caption{$cos\theta_K^{\ast}$ evolution of the differential cross sections $\sigma_U = \sigma_T + \epsilon\,\sigma_L$, $\sigma_{TT}$ and $\sigma_{LT}$, for the $K^+\Lambda$ (left) and $K^+\Sigma^0$ (right) final states, for $W = 1.875$ GeV and $Q^2 = 0.65$ GeV$^2$. The data are from Ref.~\cite{Carman06}. Line conventions as in Fig.~\ref{fig: k+_lambda_unpol} (right panel).}
\label{fig: k+_siglam_newelec}
\end{center}
\end{figure}

Without readjusting any parameter, we have confronted the RPR variants from Table~\ref{tab: kl_models} with the electroproduction data. In this way, their predictive power can be estimated. A selection of the results is contained in Figs.~\ref{fig: k+_lambda_unpol}-\ref{fig: k+_siglam_newelec}.

\begin{table}[t]
\begin{tabular*}{\columnwidth}{@{\extracolsep{\fill}} l l c c r}
\hline
& \textbf{Background} & \hspace*{-3.5mm}$\bm{D_{13}(1900)}$ & \hspace*{-2mm}$\bm{P_{11}(1900)}$ & \hspace*{-5mm}$\bm{\chi^2}$ \\ 
\hline
$\bm{K^+\Lambda}$ & & & & \\
RPR 2 & rot.$K$, rot.$K^{\ast}$
& -- & $\bigstar$ & 3.2
\\
& phase, $G^t_{K^{\ast}} < 0$  & $\bigstar$ & -- & 2.7
\\
RPR 3 & rot.$K$, cst.$K^{\ast}$ & -- & $\bigstar$ & 3.1
\\
& phase, $G^t_{K^{\ast}} > 0$ & $\bigstar$ & -- & 3.2
\\
RPR 4 & rot.$K$, cst.$K^{\ast}$ & -- & $\bigstar$ & 3.1 \\
& phase, $G^t_{K^{\ast}} < 0$  & $\bigstar$ & -- & 3.1 \\
$\bm{K^+\Sigma^0}$ & & & & \\
RPR 4$'$ & rot.$K$, cst.$K^{\ast}$ & -- & -- & 2.0 \\
& phase, $G^t_{K^{\ast}} < 0$ & & & \\ 
\hline
\end{tabular*}
\caption{RPR variants providing the best description of the $p(\gamma,K^+)\Lambda$ and $p(\gamma,K^+)\Sigma^0$ data. All models include the known $S_{11}(1650)$,  $P_{11}(1710)$, $P_{13}(1720)$ and $P_{13}(1900)$ $N^{\ast}$ states. Apart from these, each $K^+\Lambda$ variant assumes either a missing $D_{13}(1900)$ or $P_{11}(1900)$  resonance. A good description of the $K^+\Sigma^0$ channel could be achieved without the introduction of any missing resonances. The $K^+\Sigma^0$ amplitude further contains the $D_{33}(1700)$, $S_{31}(1900)$, $P_{31}(1910)$ and $P_{33}(1920)$ $\Delta^{\ast}$ states. The last column mentions $\chi^2$ in comparison with the low- and high-energy data from Refs.~\cite{Boyarski69,Vo72,Qui79,McNabb04,Bradford06,Zegers03,Sumihama05,Graal06,Lawall05}. \label{tab: kl_models}}
\end{table} 

The left panels of Fig.~\ref{fig: k+_lambda_unpol} display the $Q^2$ evolution of the unseparated ($\sigma_T+ \epsilon\,\sigma_L$) and separa\-ted ($\sigma_T$ and $\sigma_L$) $p(\mathrm{e},\mathrm{e}'K^+)\Lambda$ differential cross sections. 
The RPR variants 3 and 4 (not shown) are incompatible with these data, as they predict an unrealistically steep decrease of $\sigma_T$ as a function of $Q^2$. As seen from the figure, both RPR-2 variants describe the slope of this observable well.

Because the unpolarized electroproduction data do not allow to discriminate between the RPR-2 amplitudes assuming a missing $D_{13}$ or $P_{11}$ state, we also consider the transferred polarization for the $\overrightarrow{\mathrm{e}} p \rightarrow \mathrm{e}' K \overrightarrow{\Lambda}$ process. Figure~\ref{fig: k+_lambda_pol} compares our calculations for $P'_x$, $P'_z$, $P'_{x'}$ and $P'_{z'}$ to the data~\cite{Carman03}. 
It is clear that the RPR-2 amplitude including a $D_{13}(1900)$ state produces results far superior to those of the one assuming a $P_{11}(1900)$. Our combined analysis of the photo- and electroproduction data thus leads to the identification of RPR2 (Table~\ref{tab: kl_models}) as the preferred model variant, and of $D_{13}(1900)$ as the most likely missing-resonance candidate, in the $K^+\Lambda$ channel. 

For the $p(\mathrm{e},\mathrm{e}'K^+)\Sigma^0$ process, only unpolarized data are available. In the right panels of Fig.~\ref{fig: k+_lambda_unpol}, the separated and unseparated cross sections are compared with the results of the RPR-4$'$ variant from Table~\ref{tab: kl_models}. For all three observables, the data are well matched by the computed curves. It turns out that $\sigma_T + \epsilon \sigma_L$ and $\sigma_L$ can be reasonably well described in a pure background model, whereas reproducing the slope of $\sigma_T$ clearly requires some resonant contributions to the amplitude.

New electroproduction data from the CLAS collaboration, including measurements of the $\sigma_{LT}$ and $\sigma_{TT}$ structure functions, have recently become available~\cite{Carman06}. In Fig.~\ref{fig: k+_siglam_newelec}, we show predictions for the cos$\theta_K^{\ast}$ dependence of $\sigma_{T}+\epsilon\, \sigma_L$, $\sigma_{TT}$ and $\sigma_{LT}$ for one of the energy bins covered by CLAS, using the RPR-2 model with a missing $D_{13}$ for the $K^+\Lambda$ channel, and RPR 4$'$ for the $K^+\Sigma^0$ one. Both the full RPR-2 amplitude and its background contribution reasonably reproduce the trends of the $p(\mathrm{e},\mathrm{e}'K^+)\Lambda$ data, including the strong forward-peaking behavior of the unseparated cross section. For $p(\mathrm{e},\mathrm{e}'K^+)\Sigma^0$, the quality of agreement with the data is considerably worse. The absence of any forward peaking in this channel is, however, predicted correctly by the RPR-4$'$ model. The differences between the RPR and background results are significantly more pronounced for $\Sigma^0$ than for $\Lambda$ production, hinting that useful resonance information may be gained from the $K^+\Sigma^0$ channel. 

\section{Conclusion}
\label{sec: conclusion}

We have applied a Regge-plus-resonance (RPR) strategy, developed for kaon photoproduction from the proton~\cite{CorthalsL,CorthalsS}, to obtain a description of the $p(\mathrm{e},\mathrm{e}'K^+)\Lambda,\Sigma^0$ processes in the resonance region. It was demonstrated that the electroproduction response functions and polarization observables are particularly useful for fine-tuning certain RPR-model choices which the $p(\gamma,K)Y$ data fail to unambiguously determine. We expect that the new CLAS photo- and electroproduction data~\cite{Carman06,Schum06} will serve as a stringent test of the proposed models' predictive power, and provide important leverage for further refinements.


\end{document}